# An Energy-Efficient Mixed-Signal Neuron for Inherently Error-Resilient Neuromorphic Systems

Baibhab Chatterjee, *Student Member, IEEE*, Priyadarshini Panda, *Student Member, IEEE*,
Shovan Maity, *Student Member, IEEE*, Kaushik Roy, *Fellow, IEEE* and Shreyas Sen, *Member, IEEE*
School of Electrical and Computer Engineering, Purdue University, West Lafayette, Indiana – 47907, USA
E-mail: {bchatte, pandap, maity, kaushik, shreyas}@purdue.edu

*Abstract*—This work presents the design and analysis of a mixed-signal neuron (MS-N) for convolutional neural networks (CNN) and compares its performance with a digital neuron (Dig-N) in terms of operating frequency, power and noise. The circuit-level implementation of the MS-N in 65 nm CMOS technology exhibits 2-3 orders of magnitude better energy-efficiency over Dig-N for neuromorphic computing applications - especially at low frequencies due to the high leakage currents from many transistors in Dig-N. The inherent error-resiliency of CNN is exploited to handle the thermal and flicker noise of MS-N. A system-level analysis using a cohesive circuit-algorithmic framework on MNIST and CIFAR-10 datasets demonstrate an increase of 3% in worst-case classification error for MNIST when the integrated noise power in the bandwidth is ~ 1 μV².

## I. INTRODUCTION

The energy-efficiency of the human brain is orders of magnitude better than that of a modern-day von-Neumann computer [1], which necessitates exploring other architectures for applications with different frequency or precision requirements. Neuromorphic computing, inspired by brain, uses artificial neural networks (ANN) for applications such as classification and pattern recognition. As shown later, Dig-Ns [2-4] in an ANN offer a wide frequency of operation, but are not energy-efficient in lower frequencies due to static leakage currents of high number of transistors. On the other hand, the use of mixed-signal circuitry for computing has been a topic of debate for more than two decades [5-6], as the efficiency gained in terms of energy is often overshadowed by the effects of noise and variability. However, due to the presence of multiple distributed connections from input to output in an ANN, it offers inherent error-resiliency towards noise/variability and hence the low power consumption of MS-N can be properly leveraged.

MS-N configurations for non-learning spiking neural networks (SNN) are explored in [1, 7] that perform current-switching in large-signal mode. Convolutional neural networks (CNN), however, require a multiply and accumulate (MAC) model for existing learning algorithms (e.g. back-propagation). MAC based architectures implement summation of incoming signals through multiple synapses with different weights followed by an activation function for thresholding (Fig. 1). Though large-signal current-mode MAC structures [8] are available, a small-signal implementation with a fixed current can achieve a better power-bandwidth trade-off. In this work, we present a compact resistor-less differential MS-N with PMOS load which operate in small-signal mode, and is energy-efficient over a large range of bias currents.

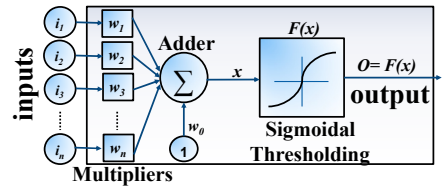

Fig. 1. Multiplier-based neuron model: MAC with thresholding

This paper is organized as follows: Section II presents the architecture of the small-signal MS-N and compares its energy-efficiency with a synthesized Dig-N. Section III describes the system-level framework on MNIST and CIFAR-10 datasets that is used to establish a relationship between the quantization/thermal/switch noise in the system and the classification error for digit/image recognition applications. Section IV concludes the paper by summarizing our key contributions.

## II. MS-N ARCHITECTURE

Fig. 2(a) shows the $N$-bit, differential-amplifier based MS-N architecture with $n$ synaptic weights. The $N$-bit weights are coming from a digital memory while the MAC operation is performed in an analog fashion, hence the name MS-N. The weights activate switches at the gate of the input transistors while a fixed bias current flows through the $k$-th synapse (for all $k = 1,2,3, … ,n$), enabling the small signal operation. Had the bias currents of the individual bits in the $k$-th synapse been controlled by the weights, the circuit would have operated in large-signal mode, hence (a) changing bandwidth with weight (b) requiring resistive loads for linearity of multiplication. Since the gain of the $k$-th branch needs to be 1 for supporting multiple synapses in the design, lower currents in the large-signal mode would have had high area penalty for the resistive load.

The output of the MS-N is modeled in Eq. (1).

$$V_{out} = \sigma(\sum g_m w_k V_k) \qquad (1)$$

where $\sigma$ is the transfer function of a differential amplifier, which acts as a sigmoidal activation function in the context of a neuron ($F$ in Fig. 1), $w_k$ and $V_k$ are the weights and the ac-coupled input voltage of the $k$-th synapse, respectively. $g_m$ is the input transconductance and $\sum g_m w_k V_k$ represent the addition of currents at the output nodes.

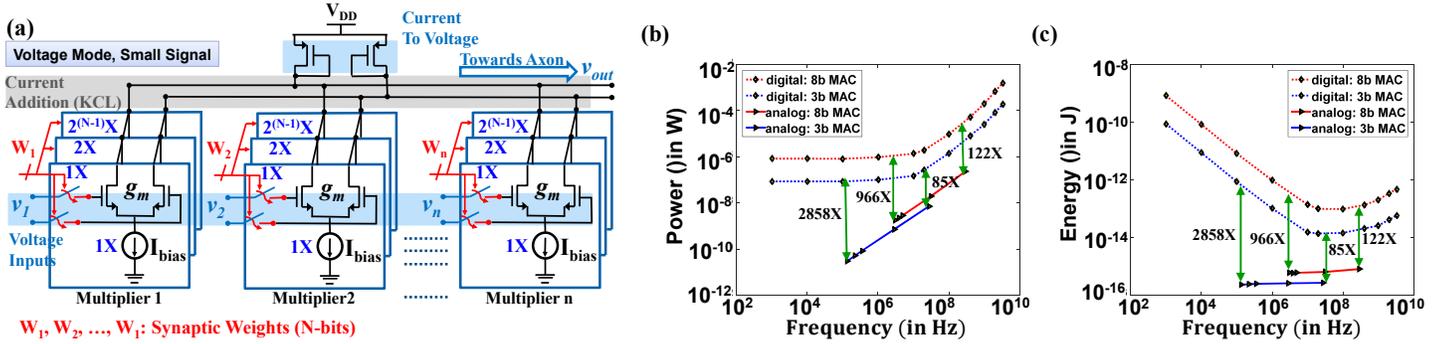

Fig. 2. (a) MS-N Architecture (N-bits, n-weights); (b) Power and (c) Energy comparison of 8-bit and 3-bit analog MAC with 8-bit and 3-bit digital MAC

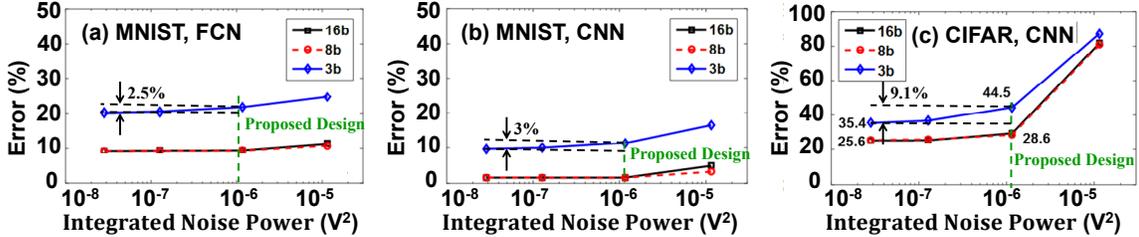

Fig. 2. System level Classification Error for 3-bit, 8-bit and 16-bit MS-N: (a) MNIST [9] with FCN, (b) MNIST with CNN and (c) CIFAR-10 [10] with CNN

## A. Power consumption with frequency

We have designed an 8-bit and a 3-bit MS-N in TSMC 65 nm CMOS technology. For comparison purpose, 8-bit and 3-bit Dig-Ns with Wallace-tree multipliers are synthesized. Fig. 2 exhibits the (b) power and (c) energy consumptions for analog and digital MACs with a single multiplier. At frequencies < 1 MHz, the all-digital implementation suffers from static leakage currents, which is quite high due to the large number of transistors (1980 MOSFETs in 8-bit MAC, 192 MOSFETs in 3-bit MAC). In contrast, power for the analog MAC scales with frequency and hence it fares extremely well at frequencies < 1 MHz, where power is about 1000 times lower than its digital counterpart. From Fig. 2 (c), it can also be inferred that the MS-N is 2-3 orders more energy-efficient as compared to the Dig-N over all frequencies.

## III. SYSTEM LEVEL EXPLOITATION OF NEUROMORPHIC ERROR-RESILIENCY FOR ENERGY-EFFICIENT MS-N

The low-power of the MS-N comes with additional thermal and flicker noise that is not present in the Dig-N. A cohesive circuit-algorithmic framework is developed to analyze the effect of this noise on ANN classification accuracy. The MNIST [9] and CIFAR-10 [10] datasets are used on a fully connected (FCN) and a convolutional (CNN) ANN to study the classification error rate for digit/image recognition problems. The noise in the bandwidth of interest is integrated using a first order rectangular approximation, and is included as a random component in the weights for MAC operation. Fig. 3 shows the performance of a 3-bit, an 8-bit and a 16-bit MS-N for (a) MNIST-FCN (b) MNIST-CNN and (c) CIFAR-10-CNN frameworks. The 8-bit and 16-bit MS-N both exhibit similar classification error across all three experiments. The worst case increase in error for MNIST is as low as 3% in CNN (3-bit) when the integrated noise power is ~ 1 $\mu V^2$, which corresponds to a unit current of 100 pA through a branch of the MS-N. The effects of variability and mismatch will be analyzed as a future work. However, such one-time manufacturing-related non-idealities may not have significant impact on the error rate if an on-chip training algorithm is employed. Another future direction of this work would be to reduce the energy of the memory-fetch stage for obtaining the synaptic weights.

## IV. CONCLUSION

Compared to a traditional Dig-N, the proposed MS-N is ~1000X more energy-efficient at low frequencies (<1 MHz), and >85X more energy-efficient across all frequencies, without significantly affecting the classification error rate for digit/image recognition problems. This is even more attractive for applications with low-frequency input signals where digital implementations require input and output FIFOs and has a trade-off between FIFO size and supply-switching of power-gated digital MACs. Energy consumption of the 8-bit MS-N is only 0.7 fJ/synapse, thus providing enough headroom for mimicking a biological neuron (20 fJ/op [1]) at the system level.

## ACKNOWLEDGMENT

The authors thank Mr. Ayan Biswas for his technical inputs and contribution towards this work.

## REFERENCES

[1] K. Boahen, "A Neuromorph's Prospectus," *IEEE Comput. Sci. Eng.*, 2017.
[2] A. Cassidy and A.G. Andreou, "Dynamical digital silicon neurons," *IEEE Biomed. Circuits and Syst. Conf.*, 2008.
[3] E. Painkras et al., "SpiNNekar: A 1-W 18-Core System-on-Chip for Massively-Parallel Neural Network Simulation", *IEEE JSSC*, 2013.
[4] F. Akopyne et al., "TrueNorth: Design and Tool Flow of a 65 mW 1 Million Neuron Programmable Neurosynaptic Chip", *IEEE Trans. Comput.-Aided Design Integr. Circuits Syst.*, 2015.
[5] R. Sarpeshkar, "Analog versus Digital: Extrapolating from Electronics to Neurobiology," *Neural Computation*, 1998.
[6] J. Hasler, "Opportunities in Physical Computing Driven by Analog Realization," *IEEE Int. Conf. on Rebooting Computing*, 2016.
[7] J. Schemmel et al., "A Wafer-Scale Neuromorphic Hardware System for Large-Scale Neural Modeling", Proc. *IEEE ISCAS.*, 2010.
[8] R. Chawla et al., "A 531 nW/MHz, 128×32 Current-Mode Programmable Analog Vector-Matrix Multiplier with Over Two Decades of Linearity," Proc. *IEEE Custom Integr. Circuits Conf.*, 2004.
[9] Y. LeCun et al., "The MNIST Database" [Online]. Available: http://yann.lecun.com/exdb/mnist/ [Accessed: 10-Jun-2017].
[10] A Krizhevsky et al., "The CIFAR-10 Dataset" [Online]. Available: https://www.cs.toronto.edu/~kriz/cifar.html [Accessed: 10-Jun-2017].